\overfullrule=0pt
\def\refto#1{$^{#1}$}           
\def\ref#1{ref.~#1}                     
\def\Ref#1{#1}                          
\gdef\refis#1{\item{#1.\ }}                    
\def\beginparmode{\endmode
  \begingroup \def\endmode{\par\endgroup}}
\let\endmode=\par
\def\body{\beginparmode}
\def\head#1{                    
  \goodbreak\vskip 0.5truein    
  {\centerline{\bf{#1}}\par}
   \nobreak\vskip 0.25truein\nobreak}
\def\references                 
Phys Rev
  {\head{References}            
   \beginparmode
   \frenchspacing \parindent=0pt \leftskip=1truecm
   \parskip=8pt plus 3pt \everypar{\hangindent=\parindent}}
\def\endreferences{\body}
\def\gs{\mathrel{\raise0.35ex\hbox{$\scriptstyle >$}\kern-0.6em
\lower0.40ex\hbox{{$\scriptstyle \sim$}}}}
\def\ls{\mathrel{\raise0.35ex\hbox{$\scriptstyle <$}\kern-0.6em
\lower0.40ex\hbox{{$\scriptstyle \sim$}}}}
\def\kms{km$\,$s$^{-1}$}

\def\apm{APM$\,$08279+5255}
\def\K{{$\,$K}}
\def\sec{{$\,$s}}

\def\COfour{CO(4$\rightarrow$3)}
\def\COnine{CO(9$\rightarrow$8)}

\catcode`@=11
\newcount\r@fcount \r@fcount=0
\newcount\r@fcurr
\immediate\newwrite\reffile
\newif\ifr@ffile\r@ffilefalse
\def\w@rnwrite#1{\ifr@ffile\immediate\write\reffile{#1}\fi\message{#1}}

\def\writer@f#1>>{}
\def\referencefile{
  \r@ffiletrue\immediate\openout\reffile=\jobname.ref%
  \def\writer@f##1>>{\ifr@ffile\immediate\write\reffile%
    {\noexpand\refis{##1} = \csname r@fnum##1\endcsname = %
     \expandafter\expandafter\expandafter\strip@t\expandafter%
     \meaning\csname r@ftext\csname r@fnum##1\endcsname\endcsname}\fi}%
  \def\strip@t##1>>{}}

\def\citeall#1{\xdef#1##1{#1{\noexpand\cite{##1}}}}
\def\cite#1{\each@rg\citer@nge{#1}}	

\def\each@rg#1#2{{\let\thecsname=#1\expandafter\first@rg#2,\end,}}
\def\first@rg#1,{\thecsname{#1}\apply@rg}	
\def\apply@rg#1,{\ifx\end#1\let\next=\relax
\else,\thecsname{#1}\let\next=\apply@rg\fi\next}

\def\citer@nge#1{\citedor@nge#1-\end-}	
\def\citer@ngeat#1\end-{#1}
\def\citedor@nge#1-#2-{\ifx\end#2\r@featspace#1 
  \else\citel@@p{#1}{#2}\citer@ngeat\fi}	
\def\citel@@p#1#2{\ifnum#1>#2{\errmessage{Reference range #1-#2\space is bad.}%
    \errhelp{If you cite a series of references by the notation M-N, then M and
    N must be integers, and N must be greater than or equal to M.}}\else%
 {\count0=#1\count1=#2\advance\count1 by1\relax\expandafter\r@fcite\the\count0,%
  \loop\advance\count0 by1\relax
    \ifnum\count0<\count1,\expandafter\r@fcite\the\count0,%
  \repeat}\fi}

\def\r@featspace#1#2 {\r@fcite#1#2,}	
\def\r@fcite#1,{\ifuncit@d{#1}
    \newr@f{#1}%
    \expandafter\gdef\csname r@ftext\number\r@fcount\endcsname%
                     {\message{Reference #1 to be supplied.}%
                      \writer@f#1>>#1 to be supplied.\par}%
 \fi%
 \csname r@fnum#1\endcsname}
\def\ifuncit@d#1{\expandafter\ifx\csname r@fnum#1\endcsname\relax}%
\def\newr@f#1{\global\advance\r@fcount by1%
    \expandafter\xdef\csname r@fnum#1\endcsname{\number\r@fcount}}

\let\r@fis=\refis			
\def\refis#1#2#3\par{\ifuncit@d{#1}
   \newr@f{#1}%
   \w@rnwrite{Reference #1=\number\r@fcount\space is not cited up to now.}\fi%
  \expandafter\gdef\csname r@ftext\csname r@fnum#1\endcsname\endcsname%
  {\writer@f#1>>#2#3\par}}

\def\ignoreuncited{
   \def\refis##1##2##3\par{\ifuncit@d{##1}%
     \else\expandafter\gdef\csname r@ftext\csname r@fnum##1\endcsname\endcsname%
     {\writer@f##1>>##2##3\par}\fi}}

\def\r@ferr{\endreferences\errmessage{I was expecting to see
\noexpand\endreferences before now;  I have inserted it here.}}
\let\r@ferences=\references
\def\references{\r@ferences\def\endmode{\r@ferr\par\endgroup}}

\let\endr@ferences=\endreferences
\def\endreferences{\r@fcurr=0
  {\loop\ifnum\r@fcurr<\r@fcount
    \advance\r@fcurr by 1\relax\expandafter\r@fis\expandafter{\number\r@fcurr}%
    \csname r@ftext\number\r@fcurr\endcsname%
  \repeat}\gdef\r@ferr{}\endr@ferences}


\let\r@fend=\endpaper\gdef\endpaper{\ifr@ffile
\immediate\write16{Cross References written on []\jobname.REF.}\fi\r@fend}

\catcode`@=12

\citeall\refto		
\citeall\ref		%
\citeall\Ref		%


\def\doublespace{\baselineskip 24pt \lineskip 10pt \parskip 5pt plus 10 pt}
\def\today{\number\day\enspace
     \ifcase\month\or January\or Febuary\or March\or April\or May\or
     June\or July\or August\or September\or October\or
     November\or December\fi \enspace\number\year}
\def\clock{\count0=\time \divide\count0 by 60
    \count1=\count0 \multiply\count1 by -60 \advance\count1 by \time
    \number\count0:\ifnum\count1<10{0\number\count1}\else\number\count1\fi}
\footline={\hss -- \folio\ -- \hss}

\def\deg{\ifmmode^\circ\else$^\circ$\fi}
\def\solar{\ifmmode_{\mathord\odot}\else$_{\mathord\odot}$\fi}
\def\jref#1 #2 #3 #4 {{\par\noindent \hangindent=3em \hangafter=1 
      \advance \rightskip by 5em #1, {\it#2}, {\bf#3}, #4.\par}}
\def\ref#1{{\par\noindent \hangindent=3em \hangafter=1 
      \advance \rightskip by 5em #1.\par}}
\newcount\eqnum
\def\nexteq{\global\advance\eqnum by1 \eqno(\number\eqnum)}
\def\lasteq#1{\if)#1[\number\eqnum]\else(\number\eqnum)\fi#1}
\def\preveq#1#2{{\advance\eqnum by-#1
    \if)#2[\number\eqnum]\else(\number\eqnum)\fi}#2}

\def\tableheight{\vrule width 0pt height 8.5pt depth 3.5pt}
{\catcode`|=\active \catcode`&=\active 
    \gdef\tabledelim{\catcode`|=\active \let|=\vbar
                     \catcode`&=\active \let&=\nobar} }
\def\table{\begingroup
    \def\twidth{\hsize}
    \def\tablewidth##1{\def\twidth{##1}}
    \def\defaultheight{\vrule width 0pt height 8.5pt depth 3.5pt}
    \def\heightdepth##1{\dimen0=##1
        \ifdim\dimen0>5pt 
            \divide\dimen0 by 2 \advance\dimen0 by 2.5pt
            \dimen1=\dimen0 \advance\dimen1 by -5pt
            \vrule width 0pt height \the\dimen0  depth \the\dimen1
        \else  \divide\dimen0 by 2
            \vrule width 0pt height \the\dimen0  depth \the\dimen0 \fi}
    \def\spacing##1{\def\defaultheight{\heightdepth{##1}}}
    \def\nextheight##1{\noalign{\gdef\tableheight{\heightdepth{##1}}}}
    \def\end{\cr\noalign{\gdef\tableheight{\defaultheight}}}
    \def\zerowidth##1{\omit\hidewidth ##1 \hidewidth}    
    \def\hline{\noalign{\hrule}}
    \def\skip##1{\noalign{\vskip##1}}
    \def\bskip##1{\noalign{\hbox to \twidth{\vrule height##1 depth 0pt \hfil
        \vrule height##1 depth 0pt}}}
    \def\header##1{\noalign{\hbox to \twidth{\hfil ##1 \unskip\hfil}}}
    \def\bheader##1{\noalign{\hbox to \twidth{\vrule\hfil ##1 
        \unskip\hfil\vrule}}}
    \def\spanloop{\span\omit \advance\mscount by -1}
    \def\extend##1##2{\omit
        \mscount=##1 \multiply\mscount by 2 \advance\mscount by -1
        \loop\ifnum\mscount>1 \spanloop\repeat \ \hfil ##2 \unskip\hfil}
    \def\vbar{&\vrule&}
    \def\nobar{&&}
    \def\hdash##1{ \noalign{ \relax \gdef\tableheight{\heightdepth{0pt}}
        \toks0={} \count0=1 \count1=0 \putout##1\end 
        \toks0=\expandafter{\the\toks0 &\end} \xdef\piggy{\the\toks0} }
        \piggy}
    \let\e=\expandafter
    \def\putspace{\ifnum\count0>1 \advance\count0 by -1
        \toks0=\e\e\e{\the\e\toks0\e&\e\multispan\e{\the\count0}\hfill} 
        \fi \count0=0 }
    \def\putrule{\ifnum\count1>0 \advance\count1 by 1
        \toks0=\e\e\e{\the\e\toks0\e&\e\multispan\e{\the\count1}\leaders\hrule\hfill}
        \fi \count1=0 }
    \def\putout##1{\ifx##1\end \putspace \putrule \let\next=\relax 
        \else \let\next=\putout
            \ifx##1- \advance\count1 by 2 \putspace
            \else    \advance\count0 by 2 \putrule \fi \fi \next}   }
\def\tablespec#1{
    \def\vdimens{\noexpand\tableheight}
    \def\tabby{\tabskip=0pt plus100pt minus100pt}
    \def\r{&################\tabby&\hfil################\unskip}
    \def\c{&################\tabby&\hfil################\unskip\hfil}
    \def\l{&################\tabby&################\unskip\hfil}
    \edef\templ{\noexpand\vdimens ########\unskip  #1 
         \unskip&########\tabskip=0pt&########\cr}
    \tabledelim
    \edef\body##1{ \vbox{
        \tabskip=0pt \offinterlineskip
        \halign to \twidth {\templ ##1}}} }

\newbox\grsign \setbox\grsign=\hbox{$>$}
\newdimen\grdimen \grdimen=\ht\grsign
\newbox\laxbox \newbox\gaxbox
\setbox\gaxbox=\hbox{\raise.5ex\hbox{$>$}\llap
	{\lower.5ex\hbox{$\sim$}}}\ht1=\grdimen\dp1=0pt
\setbox\laxbox=\hbox{\raise.5ex\hbox{$<$}\llap
	{\lower.5ex\hbox{$\sim$}}}\ht2=\grdimen\dp2=0pt

\def\uJy{\ifmmode{\,\mu{\rm Jy}}\else$\,{\mu{\rm Jy}}$\fi}
\def\mJy{\ifmmode{\,{\rm mJy}}\else${\,{\rm mJy}}$\fi}
\def\MHz{\ifmmode{\,{\rm MHz}}\else{$\,{\rm MHz}$}\fi}
\def\GHz{\ifmmode{\,{\rm GHz}}\else{$\,{\rm GHz}$}\fi}
\def\solar{\ifmmode_{\mathord\odot}\else$_{\mathord\odot}$\fi}
\def\Msolar{\ifmmode{\, {\rm M\solar}}\else{${\, {\rm M\solar}}$}\fi}
\def\Rsolar{\ifmmode{\, {\rm R\solar}}\else{${\, {\rm R\solar}}$}\fi}
\def\kms{\ifmmode{\,{\rm km\,s^{-1}}}\else${\,{\rm km\,s^{-1}}}$\fi}
\def\kpc{\ifmmode{\,{\rm kpc}}\else${\,{\rm kpc}}$\fi}
\def\us{\ifmmode{\,\mu{\rm s}}\else$\,{\mu{\rm s}}$\fi}
\def\ms{\ifmmode{\,{\rm ms}}\else$\,{{\rm ms}}$\fi}
\def\y{\ifmmode{\,{\rm y}}\else$\,{\rm y}$\fi}
\def\h{\ifmmode{^{\rm h}}\else$^{\rm h}$\fi}
\def\m{\ifmmode{^{\rm m}}\else$^{\rm m}$\fi}
\def\s{\ifmmode{^{\rm s}}\else$^{\rm s}$\fi}
\def\Lmin{\ifmmode{L_{min}}\else{$L_{min}$}\fi}

\input psfig.sty
\magnification=\magstep1
\doublespace
\font\eightrm=cmr8
\font\lgh=cmbx10 scaled \magstep2
\def\hb{\hfill\break}



\noindent{\hfill Version: \today, I09100 LS/ks}
\smallskip
\hrule
\bigskip
\line{\lgh A massive reservoir of low-excitation molecular gas \hb}
\line{\lgh at high redshift \hb}

\bigskip
\bigskip

\line{Padeli Papadopoulos,$^*$
      Rob Ivison,$^{\dagger}$ Chris Carilli$^{\ddagger}$
      and Geraint Lewis$^{\parallel}$      \hb}

\bigskip

\line{$^*$
      Leiden Observatory, PO Box 9513, 2300 Leiden, The Netherlands \hfill}

\line{$^\dagger$
      Department of Physics and Astronomy, University College London,
      Gower Street, \hfill}
\line{$\;\,$ London WC1E 6BT, UK \hfill}

\line{$^\ddagger$
      National Radio Astronomy Observatories, PO Box 0,
                    Socorro, NM 87801-0387, USA \hfill}

\line{$^\parallel$
      Anglo-Australian Observatory, PO Box 296, Epping,
                    NSW 1710, New South Wales, Australia \hfill}

\bigskip
\hrule
\bigskip

\noindent{\bf
Molecular hydrogen (H$_2$) is an important component of galaxies
because it fuels star formation and accretion onto active galactic
nuclei (AGN), the two processes that generate the large infrared
luminosities of gas-rich galaxies\refto{T88,SSS91}. Observations of
spectral-line emission from the tracer molecule carbon monoxide (CO)
are used to probe the properties of this gas. But the lines that have
been studied in the local Universe --- mostly the lower rotational
transitions of $J = 1 \rightarrow 0$ and $J = 2 \rightarrow 1$ ---
have hitherto been unobservable in high-redshift galaxies. Instead,
higher transitions have been used, although the densities and
temperatures required to excite these higher transitions may not be
reached by much of the gas. As a result, past observations may have
underestimated the total amount of molecular gas by a substantial
amount. Here we report the discovery of large amounts of
low-excitation molecular gas around the infrared-luminous quasar,
\apm\ at $z$ = 3.91, using the two lowest excitation lines of
$^{12}$CO ($J = 1 \rightarrow0$ and $J = 2 \rightarrow 1$). The maps
confirm the presence of hot and dense gas near the
nucleus\refto{De99}, and reveal an extended reservoir of molecular gas
with low excitation that is 10 to 100 times more massive than the gas
traced by higher-excitation observations. This raises the possibility
that significant amounts of low-excitation molecular gas may lurk in
the environments of high-redshift ($z>3$) galaxies.

}

\bigskip


\apm\ was discovered serendiptously during a search for Galactic halo
carbon stars\refto{Ie98}.  It comprises a strongly lensed broad
absorption line quasar at $z$ = 3.9 with a complex optical spectrum
and a bolometric luminosity of $\sim10^{14}\,$L$_{\odot}$ (where
L$_{\odot}$ is the luminosity of the Sun) after correcting for
lensing magnification; it displays extremely strong rest-frame
far-infrared emission\refto{Le98} suggestive of dust-rich gas
undergoing rapid star formation.  The \COfour\ and \COnine\ lines have
been detected\refto{De99} suggesting that the molecular gas must be
warm, extended on a scale of up to $1''$ in the image plane (only
160--270$\,$pc in the source plane), with a mass of $(1-6) \times
10^9\,$M$_{\odot}$.

We have used the National Radio Astronomy Observatory's Very Large
Array (VLA) to detect and image CO $J$=1$\rightarrow$0 and
$J$=2$\rightarrow$1 emission from \apm. Emission in the
CO(1$\rightarrow$0) line (Fig.\ 1) extends over scales of $\sim7''$
($\sim 30\,$kpc at $z$ = 3.9 for $H_0 =
75\,$km\sec$^{-1}\,$Mpc$^{-1}$, $q_0=0.5$), well beyond the
gravitationally lensed nucleus\refto{Ie99} where the emission from the
CO(4$\rightarrow$3) and CO(9$\rightarrow$8) transitions is
confined\refto{De99}. Maps of CO(2$\rightarrow$1) detect the northern
part of the extended emission and show it to consist of two
kinematically and spatially distinct features (Fig.\ 2).  The
continuum emission detected at 23$\,$GHz emanates from the lensed
nucleus and is unresolved ($\leq 0.6''$) with a flux density of
$0.30\pm 0.05\,$mJy.  At 43$\,$GHz, no continuum emission is detected
at a level of $\sigma = 0.30$ mJy, and the emission expected from the
hot dust in the nucleus\refto{Le98} is negligible ($\leq 0.04\,$mJy).
No appreciable dust continuum is expected underlying the extra-nuclear
CO(2$\rightarrow$1) emission from the cooler dust/gas in those
regions.

The nuclear CO(1$\rightarrow$0) line emission has a
velocity-integrated flux density of $0.150\pm0.045\,$Jy\kms\ and a
deconvolved size of $\sim 0.6''-1''$, confirmed by the
CO(2$\rightarrow$1) map (Fig.\ 2) which shows this emission extending
$\sim 0.8''$ along the direction expected from matched-resolution
1.6-$\mu$m Hubble Space Telescope (HST) images\refto{Ie99}.  This size
is similar to that of the CO(4$\rightarrow$3) and CO(9$\rightarrow$8)
emission\refto{De99}, while the velocity-averaged brightness
temperature ratios, $R_{4\rightarrow 3}=(4\rightarrow 3)/(1\rightarrow
0)$ and $R_{2\rightarrow 1}=(2\rightarrow 1)/(1\rightarrow 0)$, are
$1.5\pm 0.5$ and $1.35\pm0.55$, respectively, values perfectly
compatible with a hot and dense gas phase in the nucleus.

We have considered in detail, and rejected, gravitational lensing as
the cause of the extended CO(2$\rightarrow$1) structure. There are no
obvious optical counterparts and the CO(2$\rightarrow$1) maps
demonstrate that the two features to the north and north-east and the
quasar nucleus do not share a common CO(2$\rightarrow$1) line profile,
a necessary characteristic if a single galaxy underlies them
all. Detailed modeling of the lens\refto{Ie99,Ee00}, based on
high-resolution HST and Keck data, does not suggest the
existence of additional lensed features.  We thus conclude that the
extra-nuclear features are due to companion galaxies near
\apm. Moreover, their large angular distance from any central lens
caustics precludes significant magnification (employing an elliptical
mass distribution to represent the lensing galaxy\refto{Ie99}, which
accurately recovers the quasar image configuration, the magnification
in this region is $\le$1.2) and hence the implied molecular gas masses
are large.

$X_{\rm CO}$ is the so-called CO-to-H$_2$ conversion
factor\refto{YS82,YS91,DSS86}, ranging from $\sim 1\,$M$_{\odot}$ (K
\kms\ pc$^2$)$^{-1}$ in ultraluminous infrared galaxies\refto{DS98}
(ULIRGs) to $\sim 5\,$M$_{\odot}$ (K \kms\ pc$^2$)$^{-1}$ for
quiescent spirals. The integrated flux density of the two
extra-nuclear features in the CO(2$\rightarrow$1) maps is
$1.15\pm0.54\,$Jy$\,$\kms\ which (for the aforementioned range of
$X_{\rm CO}$ and $R_{2\rightarrow 1}\sim 1$) yields a mass of $M({\rm
H}_2)\sim (0.65-3.2)\times 10^{11}\,$M$_{\odot}$, this value being
merely a lower limit if sub-thermal excitation of CO(2$\rightarrow$1)
($R_{2\rightarrow 1}<1$) or sub-solar metallicities ($X_{\rm
CO}>5\,$M$_{\odot}$ (K \kms\ pc$^2$)$^{-1}$) are important.

This molecular gas reservoir is $\sim 10-100$ times more massive than
that deduced for the nuclear region\refto{De99} and neither
CO(4$\rightarrow$3), CO(9$\rightarrow$8) nor high-resolution 210-GHz
continuum maps give any hint of it. This mass is comparable to that of
an entire spiral galaxy and is confined within $2''-3''$ or
$\sim10\,$kpc (Fig.~2). Similar gas-rich companions with no obvious
optical counterparts, akin to the recently discovered submillimetre
galaxy population\refto{SIB}, have been found in the vicinity of other
high-redshift systems\refto{Ie00,Pe00}. They have the tell-tale marks
of intense ($L_{\rm FIR}\sim 10^{13}\,$L$_{\odot}$), heavily obscured
starbursts in merger configurations, converting a large reservoir of
molecular gas ($\sim 10^{11}\,$M$_{\odot}$) into stars. In these
systems the dynamical mass is comparable to the gas mass, a situation
possible for the companion galaxies to \apm, though only observations
with higher spatial resolution and wider velocity coverage can address
this issue properly.
 
The failure of sensitive spectral-line observations to detect any
extra-nuclear CO(4$\rightarrow$3) and CO(9$\rightarrow$8) emission
implies low values ($\leq 0.15$) for the global $R_{4\rightarrow 3}$
and $R_{9\rightarrow 8}$ ratios for the massive reservoirs of
molecular gas near \apm. This has far-reaching consequences: at $z>3$,
millimetre-wave telescopes can only access CO $J+1 \rightarrow J$,
$J>2$ and hence cannot hope to trace the bulk of the H$_2$ gas.

The low excitation of lines such as CO(9$\rightarrow$8) is hardly
surprising, even for a starburst environment, since the {\it average}
conditions in the molecular gas will not be sufficient for their
large-scale excitation. If the physical conditions in
Orion~A\refto{Ple00} are typical of a starburst environment,
$R_{9\rightarrow 8}\leq 0.15$ (and in most cases $R_{9\rightarrow
8}\leq 0.05$).  Moreover, the molecular interstellar medium in intense
starburst environments may not be just an ensemble of warm and dense
Orion-type clouds. Where the most energetic star formation takes
place, the gas may also have a warm but diffuse
phase\refto{DS98,Ae95}, $n({\rm H}_2)\leq 10^3\,$cm$^{-3}$, generated
by the disruptive effects of intense ultraviolet radiation, supernovae
and the violent dynamics associated with mergers. The CO $J+1
\rightarrow J$, $J>2$ emission from such gas will be weak, possibly
for the entire starburst if this phase is dominant.  Starbursts with
similar far-infrared and low-$J$ CO luminosities can be dominated by
drastically different gas phases\refto{Je95,PJI97}.  Such differences
are expected to occur during the evolution of a starburst, depending
on the complex interplay between the accumulation of dense gas that
feeds the star formation (for example, through a merger) and the cloud
disruption caused by the products of that process, the massive stars.

Finally, since CO observations at high redshift usually sample
emission averaged over several kiloparsecs, the possibility of cold
and/or diffuse gas dominating the {\it global} CO emission should not
be discounted.  The CO $J+1 \rightarrow J$, $J>2$ transitions may then
have low excitation and thus be very faint, a state not altered by the
enhanced cosmic microwave background\refto{Pe00}. The detection of CO
at $z>3$ means that significant metal-enrichment has taken place;
low-excitation gas would then represent a previously metal-enriched
part of its interstellar medium that is not participating in the
observed star-formation episode.  Such excitation gradients have been
found in several nearby infrared-luminous galaxies hosting powerful
circum-nuclear starbursts and AGN\refto{PA00}.

\apm\ can be seen as an extreme case of such an excitation
gradient, further exaggerated by gravitational lensing of a
particularly hot region of the underlying system.  Differential
lensing seems to be preferentially boosting the far-infrared and CO
emission of the AGN-heated (rather than the starburst-heated)
component of the interstellar medium.  This is strongly suggested by
the high temperature of the gas and dust ($\sim 200$\K) which is not
typical of nearby starbursts\refto{HGR94} or clouds heated by young
stars (e.g.\ Orion~A, $T\sim 40-60$\K). Also, the ratio $L_{\rm
FIR}/M({\rm H}_2)\sim 10^4\,$L$_{\odot}\,$M$^{-1}_{\odot}$ is $\sim$40
times higher than the highest values observed in ULIRGs and the
$L_{\rm FIR}/L_{\rm 1.4\,GHz}$ ratio is an order of magnitude larger
than in star-forming galaxies. This all suggests a heating source
other than stars and the AGN is the obvious alternative.  Differential
magnification of AGN-heated gas therefore seems to be responsible for
the high luminosity of the CO(9$\rightarrow$8) line.

While highly excited regions with CO emission amplified by
gravitational lensing are good markers of the presence of molecular
gas at high redshifts, they may give a poor representation of its
average physical conditions, particularly of its total mass and
distribution.  At $z>3$, this excitation bias becomes more serious
since millimetre-wave interferometers, the instruments usually used
for such searches, can detect only CO $J+1 \rightarrow J$,~$J>2$ whose
excitation requirements are $n({\rm H}_2)\geq 10^4\,$cm$^{-3}$ and
$T\geq 50$\K. These values, being typical of the average conditions in
star-forming clouds, mark a gradual excitation turnover.

Observations of the CO(1$\rightarrow$0) and CO(2$\rightarrow$1)
transitions, with their minimal excitation requirements, may reveal
much larger molecular gas reservoirs at high redshifts. The VLA is
currently the only instrument capable of sensitive, sub-arcsecond
observations of these lines and will thus be an important tool for
unbiased surveys of metal-enriched H$_2$ gas around objects in the
distant Universe\refto{Ie96}.

\bigskip
\noindent{\bf Methods.}

Our observations took place during 23--24 April 2000 using the VLA in
its C configuration. After all overheads, 20 hours were spent
integrating on \apm. 19 of the 27 antennas were equipped with 43-GHz
receivers ($T_{\rm sys}$ = 45--100\K); all were equipped with 22-GHz
receivers, including eleven new receivers with $T_{\rm sys}\sim 40$\K\
(c.f.\ 100--150\K\ for the remainder).

We exploited the new fast-switching technique\refto{CH99,CMY99},
recording data every 3.3\sec, with 30\sec\ on the calibrator and
170\sec\ on the source (210\sec\ in the 22-GHz band); the typical
slewing overhead to the compact phase calibrator, 0824+558,
3.3$^{\circ}$ away, was 7\sec.  Pointing accuracy was checked every
hour.

This technique yields diffraction-limited images at the highest VLA
operating frequencies over long baselines. Conventional phase
referencing would not have been able to track tropospheric phase
variations and self-calibration techniques could not be employed
because of the low signal-to-noise per baseline.

We used continuum mode, placing emphasis on sensitivity rather than
line profile information.  A dual-polarisation 50-MHz band was placed
as close to the expected line centre as could be allowed by correlator
limitations: 23.4649$\,$GHz for CO $J$=1$\rightarrow$0 ($+91\,$\kms\
from the line centre\refto{De99}). Bandpass roll-off limited the
effective bandwidth per IF to $\sim$45$\,$MHz which, at $z$ = 3.9,
corresponds to $\Delta v\sim 575\,$\kms\ at 23$\,$GHz.  The remaining
IF pair was tuned to {\it simultaneously} observe the continuum at
23.3649$\,$GHz ($+1280\,$\kms).

We obtained matching velocity coverage at 43$\,$GHz (CO
$J$=2$\rightarrow$1) by placing two contiguous 50-MHz
dual-polarisation bands at 46.9399$\,$GHz ($-20\,$\kms\ from the line
centre).  The continuum was observed on a separate occasion with both
IF pairs tuned to 43.3$\,$GHz. The flux density scale was fixed using
3C286; the uncertainty is $\sim$15\% at 23$\,$GHz and $\sim$20\% at
46$\,$GHz. Calibration and reduction of the data was standard in most
respects and the rms noise in all the maps is similar to the expected
theoretical limit.

We cannot completely rule out the possibility that phase errors ---
caused by atmospheric fluctuations too fast to be tracked even by our
fast-switching scheme --- are the cause of the extended CO emission,
but we consider it remote. This is borne out by an extensive series of
tests: inspection of the raw visibilities; separate imaging of left-
and right-hand polarization maps; imaging of the most phase-stable
subset of the data; imaging of the calibrator source. The agreement
between the nuclear CO $J=2-1$ emission and its associated 3.5-cm
continuum is also strong testimony to the coherence of the CO $J=2-1$
map; the point-like 23.4-GHz continuum emission, observed
simultaneously with the neighbouring line emission, provides similar
supporting evidence for the resolved CO $J=1-0$ emission.

\vfill\eject


\def\araa{Ann. Rev. Astron. \& Astrophys.}
\def\apj{Astrophys.\ J.}
\def\apjl{Astrophys.\ J.}
\def\aj{Astron.\ J.}
\def\mnras{Mon.\ Not.\ R.\ Astron.\ Soc.}
\def\aap{Astron. \& Astrophys.}

\bigskip

\noindent
{\bf References.}

\refis{Ae95} 
       Aalto, S., Booth, R.\ S., Black, J.\ H.\ \& Johansson, L.\ E.\ B.
       Molecular gas in starburst galaxies: line intensities and physical 
       conditions.
       {\it \aap}\ {\bf 300,} 369--384 (1995).

\refis{CH99}
       Carilli, C.\ L.\ \& Holdaway, M.\ A.
       Tropospheric phase calibration in millimeter interferometry.
       {\it Radio Science} {\bf 34,} 817--840 (1999).

\refis{CMY99}
       Carilli, C.\ L., Menten, K.\ M.\ \& Yun, M.\ S.
       Detection of CO(2--1) and radio continuum emission from the z = 4.4
       QSO BRI$\,$1335$-$0417.
       {\it \apjl}\ {\bf 521,} L25--L28 (1999).

\refis{DSS86} 
	Dickman, R.\ L., Snell, R.\ L.\ \& Schloerb, F.\ P.
        Carbon monoxide as an extragalactic mass tracer.
        {\it \apj}\ {\bf 309,} 326--330 (1986).

\refis{DS98} 
       Downes, D.\ \& Solomon, P.\ M.
       Rotating nuclear rings and extreme starbursts in ultraluminous
       galaxies.
       {\it \apj}\ {\bf 507,} 615--654 (1998).

\refis{De99}
       Downes, D., Neri, R., Wiklind, T., Wilner, D.\ J.\ \& Shaver, P.\ A.
       Detection of CO(4--3), CO(9--8), and dust emission in the broad
       absorption line quasar APM$\,$08279+5255 at a redshift of 3.9.
       {\it Astrophys.\ J.}\ {\bf 513,} L1--L4 (1999).

\refis{Ee00}
       Egami, E.\ {\it et al.}
       APM$\,$08279+5255: Keck near- and mid-infrared high-resolution imaging.
       {\it \apj}\ {\bf 535,} 561--574 (2000).

\refis{HGR94} 
       Hughes, D.\ H., Gear, W.\ K.\ \& Robson, E.\ I.
       The submillimetre structure of the starburst nucleus in M82 --- a 
       diffraction-limited 450-micron map.
       {\it \mnras}\ {\bf 270,} 641--649 (1994).

\refis{Ie99} 
       Ibata, R.\ A., Lewis, G.\ F., Irwin, M.\ J., Leh{\'a}r, J.\ \& 
       Totten, E.\ J.
       NICMOS and VLA observations of the gravitationally lensed ultraluminous
       BAL quasar APM$\,$08279+5255: detection of a third image.
       {\it \aj}\ {\bf 118,} 1922--1930 (1999).

\refis{Ie98}
       Irwin, M.\ J., Ibata, R.\ A., Lewis, G.\ F.\ \& Totten, E.\ J.
       APM$\,$08279+5255: an ultraluminous broad absorption line quasar at a
       redshift z = 3.87.
       {\it Astrophys.\ J.}\ {\bf 505,} 529--535 (1998).

\refis{Ie96}
       Ivison, R.\ J., Papadopoulos, P.\ P., Seaquist, E.\ R.\ \& Eales,
       S.\ A.
       A search for molecular gas in a high-redshift radio galaxy.
       {\it Mon.\ Not.\ R.\ Astron.\ Soc.}\ {\bf 278,} 669--672 (1996).

\refis{Ie00}
       Ivison, R.\ J.\ {\it et al.}
       An excess of submillimeter sources near 4C$\,$41.17: a candidate
       proto-cluster at z=3.8?
       {\it Astrophys.\ J.}\ {\bf 542,} 27--34 (2000).

\refis{Je95} 
	Jackson, J.\ M., Paglione, T.\ A.\ D., Carlstrom, J.\ E.\ \& 
	Rieu, N.
        Submillimeter HCN and HCO+ emission from galaxies.  
	{\it \apj}\ {\bf 438,} 695--701 (1995).

\refis{Le98}
       Lewis, G.\ F., Chapman, S.\ C., Ibata, R.\ A., Irwin, M.\ J.\ \&
       Totten, E.\ J.
       Submillimeter observations of the ultraluminous broad absorption line
       quasar APM$\,$08279+5255.
       {\it Astrophys.\ J.}\ {\bf 505}, L1--L5 (1998).

\refis{PJI97} 
	Paglione, T.\ A.\ D., Jackson, J.\ M.\ \& Ishizuki, S.
	The average properties of the dense molecular gas in galaxies.
	{\it \apj}\ {\bf 484,} 656--663 (1997).

\refis{PA00} 
	Papadopoulos, P.\ P.\ \& Allen, M.\ L.
        Gas and dust in NGC 7469: submillimeter imaging and CO $J$=3--2.
        {\it \apj}\ {\bf 537,} 631--637 (2000).

\refis{Pe00} 
	Papadopoulos, P.\ P.\ {\it et al.}
        CO(4-3) and dust emission in two powerful high-z radio galaxies,
        and CO lines at high redshifts.  
	{\it \apj}\ {\bf 528,} 626--636 (2000).

\refis{Ple00}
       Plume, R.\ {\it et al.} 
       Large-scale $^{13}$CO $J$=5--4 and [C$\,$I] mapping of Orion A.
       {\it \apj}\ {\bf 539,} L133--L136 (2000).

\refis{SSS91} 
       Sanders, D.\ B., Scoville, N.\ Z.\ \& Soifer, B.\ T.
       Molecular gas in luminous infrared galaxies.
       {\it \apj}\ {\bf 370,} 158--171 (1991).

\refis{SIB}
       Smail, I., Ivison, R.\ J.\ \& Blain, A.\ W.
       A deep sub-millimeter survey of lensing clusters: a
       new window on galaxy formation and evolution.
       {\it Astrophys.\ J.}\ {\bf 490,} L5--L8 (1997).

\refis{T88} 
       Telesco, C.\ M.
       Enhanced star formation and infrared emission in the centers
       of galaxies.
       {\it \araa}\ {\bf 26,} 343--376 (1988).

\refis{YS82} 
       Young, J.\ S.\ \& Scoville, N.\ Z.
       Extragalactic CO --- gas distributions which follow the light in
       IC 342 and NGC 6946.
       {\it \apj}\ {\bf 258,} 467--489 (1982).

\refis{YS91} 
       Young, J.\ S.\ \& Scoville, N.\ Z.
       Molecular gas in galaxies.
       {\it \araa}\ {\bf 29, } 581--625 (1991).

\endreferences

\bigskip
\noindent
{\bf Acknowledgments.}

\noindent
PP and RI would like to acknowledge many useful discussions and the
early encouragement of Dr Ernie Seaquist. We are also indebted to Dr
Richard Barvainis.  The National Radio Astronomy Observatory is a
facility of the National Science Foundation, operated under
cooperative agreement by Associated Universities, Inc.

\bigskip
\noindent
Correspondence and requests for materials should be addressed to Rob
Ivison (rji@star.ucl.ac.uk).

\vfill\eject

\noindent{\bf Figure Captions.}

{\eightrm
\noindent {\bf Figure 1.} Molecular gas in and around \apm,
as traced by CO $J$=1$\rightarrow$0, and continuum emission from the active
nucleus. {\it Top:} Naturally-weighted maps at $\nu_{\rm obs}$ =
23.4649$\,$GHz, {\it middle:} $\nu_{\rm obs}$ = 23.3649$\,$GHz,
corresponding to CO $J$=1$\rightarrow$0 ($\nu_{\rm rest}$ =
115.2712$\,$GHz) and its adjacent continuum, respectively, at
$z=3.912$. The maps were produced from the Fourier transform of the
measured visibilities, tapered to enhance low brightness extended
emission. The dirty beam, identical in both maps with FWHM $1.5''
\times 1.4''$, is shown bottom left. Contours are at $-3, -2, 2, 3, 4,
5, 6, 8, 10 \times \sigma$, where $\sigma = 40\,\mu$Jy$\,$beam$^{-1}$.
{\it Bottom:} Naturally-weighted map of CO $J$=1$\rightarrow$0 with the
continuum source subtracted. The subtraction was performed in the
visibility domain and the resulting $uv$ data were then Fourier
transformed, CLEANed and convolved to a circular beam, FWHM $2.25''$,
shown bottom left. Contours this time are at $-3, -2, 2, 3, 4, 5, 6
\times \sigma$. The cross marks the position of lens\refto{Ie99} component
`A' at RA 08h 31min 41.64s and Dec.\ +52$^{\circ}$ 45$'$ 17.5$''$
(J2000). The 0.5$''$ offset between the peak radio intensity and `A'
is most likely due to {\it HST} astrometric errors.}

\bigskip

{\eightrm
\noindent {\bf Figure 2.} Molecular gas in and around \apm,
as traced by CO $J=2-1$, overlaid on a greyscale image of 8.45-GHz
continuum emission from the active nucleus. {\it Top:}
Naturally-weighted map of CO $J$=2$\rightarrow$1 ($\nu_{\rm rest}$ =
230.5380$\,$GHz) with both IF bands averaged (contours) and overlaid
on continuum emission at 8.45$\,$GHz (greyscale) at a common
resolution ($0.5''\times~0.5''$).  Contours are at $-3, -2, 2, 3, 4, 5
\times \sigma$, where $\sigma = 0.15\,$mJy$\,$beam$^{-1}$ and the
greyscale range is 25($2\sigma$)--305$\,\mu$Jy$\,$beam$^{-1}$. CLEAN
was applied. {\it Bottom:} Naturally-weighted, tapered maps of CO
$J$=2$\rightarrow$1 for the two IF bands separately, at a common
resolution of $0.90''\times~0.75''$ FWHM. Contours are at $-3, 3, 4,
5, 6\ \times \sigma$, where $\sigma = 0.20\,$mJy$\,$beam$^{-1}$. CLEAN
was not applied. In all cases, the beam FWHM is shown bottom left and
the cross marks the position of lens\refto{Ie99} component `A'.}

\vfill
\eject

\centerline{\hbox{\psfig{file=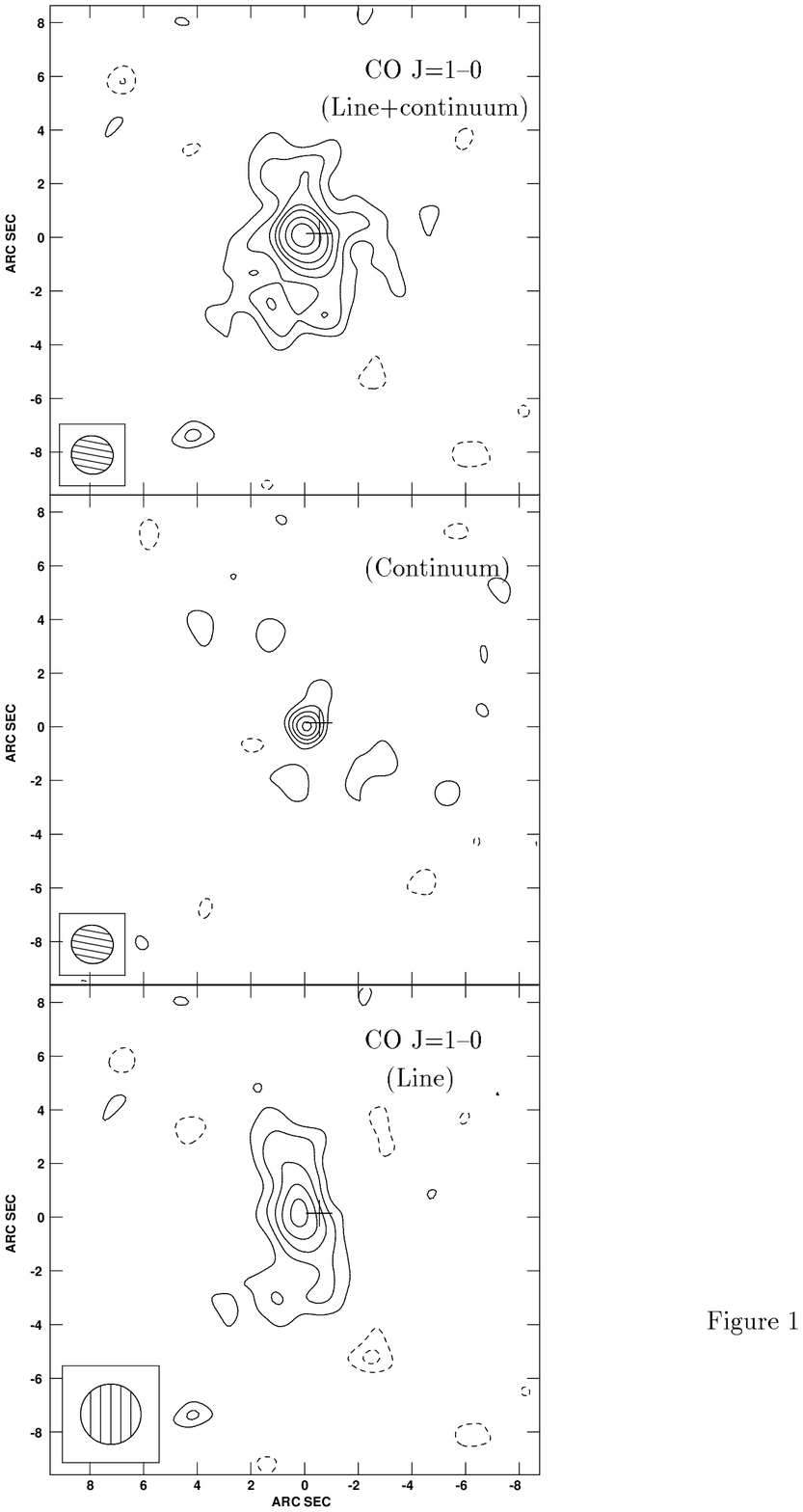,width=5.8in,angle=0}}}

\vfill
\eject

\centerline{\hbox{\psfig{file=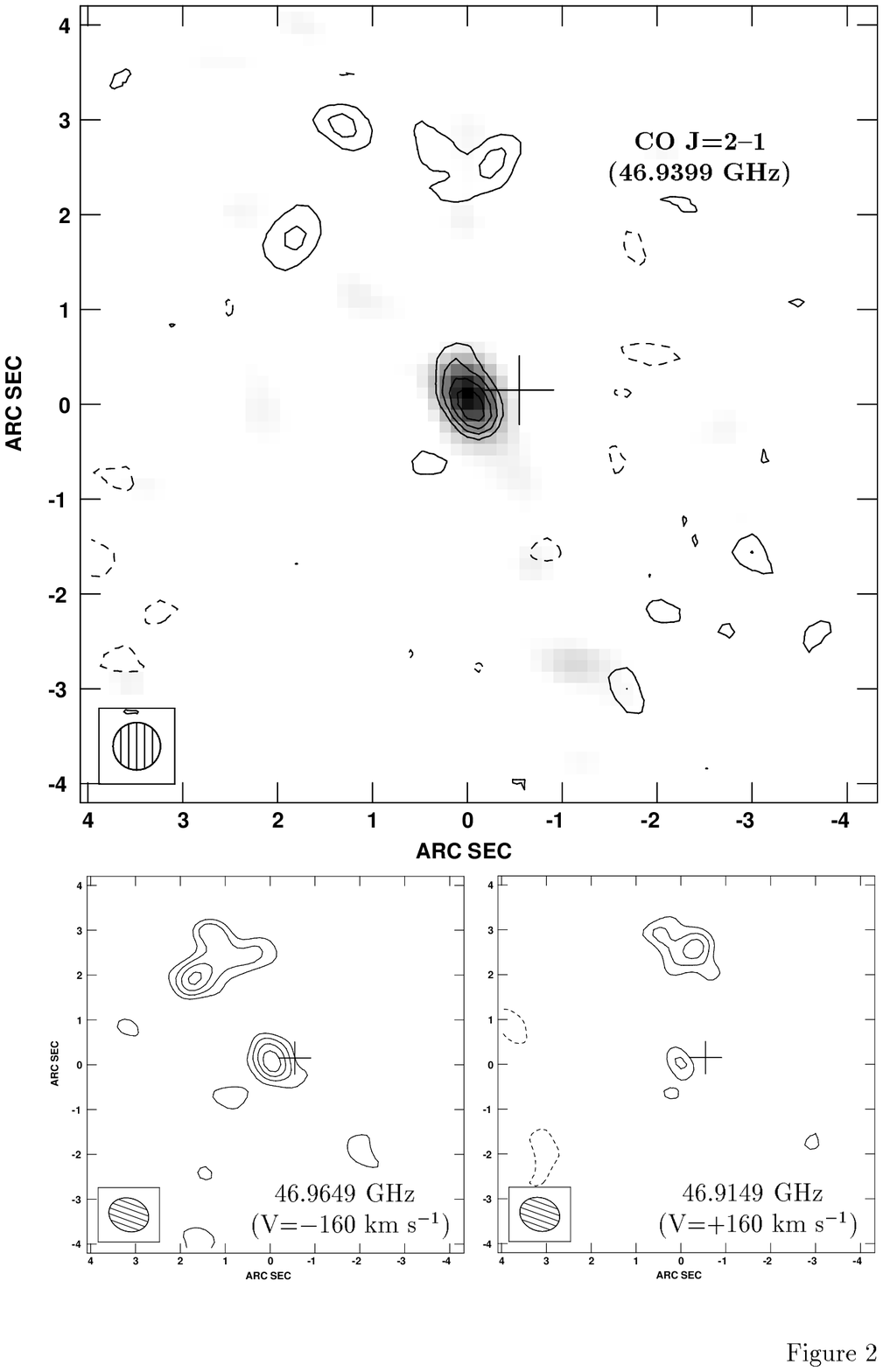,width=5.8in,angle=0}}}

\vfill
\eject

\bye